\documentclass[12pt]{article}
\usepackage{amssymb,amsmath,epsfig}

\begin{document}
\title{\bf The Stability of a Shearing Viscous Star with Electromagnetic Field}

\author{M. Sharif$^1$ \thanks{msharif.math@pu.edu.pk} and M. Azam$^{1,2}$
\thanks{azammath@gmail.com}\\
$^1$ Department of Mathematics, University of the Punjab,\\
Quaid-e-Azam Campus, Lahore-54590, Pakistan.\\
$^2$ Division of Science and Technology, University of Education,\\
Township Campus, Lahore-54590, Pakistan.}

\date{}

\maketitle
\begin{abstract}
We analyze the role of electromagnetic field for the stability of
shearing viscous star with spherical symmetry. Matching conditions
are given for the interior and exterior metrics. We use perturbation
scheme to construct the collapse equation. The range of instability
is explored in Newtonian and post Newtonian (pN) limits. We conclude
that the electromagnetic field diminishes the effects of shearing
viscosity in the instability range and makes the system more
unstable at both Newtonian and post Newtonian approximations.
\end{abstract}

\section{Introduction}

Charged self-gravitating objects have the tendency of undergoing
many phases during gravitational collapse resulting a charged black
holes or naked singularities \cite{1}. The stability of these exact
solutions is an interesting subject under the perturbation scheme.
It is believed that these stars with huge charge cannot be stable
\cite{2,3}. However, the electric charge has great relevance during
the structure formation and evolution of astrophysical objects
\cite{4}-\cite{11}.

A stellar model may be stable in one phase and later becomes
unstable in another phase. The stability of this model is subjected
against any disturbance. Dynamical instability of self-gravitating
objects is interrelated with structure as well as evolution of
astrophysical objects. In this scenario, Chandrasekhar \cite{12}
investigated the problem of dynamical instability for the isotropic
perfect fluid of a pulsating system and found the instability range
in terms of adiabatic index $\Gamma<\frac{4}{3}$.

It is well discussed in literature that the instability range of the
fluid would be decreased or increased through different physical
properties of the fluid. In this context, the dynamical instability
for adiabatic, non-adiabatic, anisotropic and shearing viscous
fluids have been explored \cite{13}-\cite{17}. Chan \cite{18} found
that both pressure anisotropy and effective adiabatic index are
increased by the shearing viscosity in a collapsing radiating star.
Horvat et al. \cite{19} have used the quasi-local equation of state
\cite{20,21} to explore the stability of anisotropic stars under
radial perturbations. Sharif and Kausar \cite{22} investigated
stability of expansion-free fluid in $f(R)$ gravity and found that
stability of the fluid is constrained by energy density
inhomogeneity, pressure anisotropy and $f(R)$ model.

The study of charged self-gravitating objects in the context of
coupled Einstein-Maxwell field equations leads to the evolution of
black hole \cite{23}-\cite{25}. The physical aspects of the
electromagnetic field has a significant role in general relativity.
Stettner \cite{26} investigated the stability of a pulsating sphere
with a constant surface charge. Glazer \cite{27} generalized this
result by taking an arbitrary distribution of charge and found that
the Bonner's charged dust model is dynamically unstable. The
stability limit for charged spheres has been proposed by many people
\cite{28}-\cite{31} starting from the Buchdahl \cite{32} work for
neutral spheres. Recently, we have investigated the problem of
dynamical instability of cylindrical and spherical systems with
electromagnetic field at Newtonian and pN regimes \cite{33}.

Here we explore the role of electromagnetic field in the stability
of collapsing fluid undergoing dissipation with shearing viscosity.
Darmois matching conditions \cite{34} are formulated for the
continuity of interior general spherically symmetric solution to the
exterior vacuum Reissner-Nordstr$\ddot{o}$m solution. The paper is
planned as: In the next section, we discuss some basic properties of
the viscous fluid, Einstein-Maxwell equations and junction
conditions. Section \textbf{3} provides the perturbation scheme to
form the collapse equation. In section \textbf{4}, we explore the
collapse equation at Newtonian and pN regimes. Finally, we discuss
our conclusion in section \textbf{5}.

\section{Fluid Distribution, Field Equations and Junction Conditions}

We consider a timelike three-space spherical surface, $\Sigma$,
which separates the $4D$ geometries into two regions interior
$M^{-}$ and exterior $M^{+}$. The $M^{-}$ is given by the general
spherically symmetric spacetime in the comoving coordinates
\begin{equation}\label{1}
ds^2_-=-A^2(t,r)dt^{2}+B^2(t,r)dr^{2}+R^2(t,r)(d\theta^{2}
+\sin^2\theta{d\phi^2}).
\end{equation}
For the exterior metric, we take Reissner-Nordstr$\ddot{o}$m metric
describing the radiation field around a charged spherically
symmetric source of the gravitational field
\begin{equation}\label{2}
ds^2_+=-\left(1-\frac{2M}{r}+\frac{Q^2}{r^2}\right)d\nu^2
-2d{\nu}d{r}+r^2(d\theta^2+\sin^2{\theta}d\phi^2),
\end{equation}
where $M$ and $Q$ are the total mass and charge respectively. The
fluid under consideration is locally dissipative in the form of
shearing viscosity and the interior energy-momentum tensor of
charged dissipative fluid is given by
\begin{equation}\label{3}
T^-_{\alpha\beta}=(\rho+\hat{p}-\xi{\Theta})u_{\alpha}u_{\beta}+(\hat{p}-\xi{\Theta})g_{\alpha\beta}
-\eta{\sigma_{{\alpha}{\beta}}}+\frac{1}{4\pi}\left(F^\gamma_{\alpha}F_{\beta\gamma}
-\frac{1}{4}g_{\alpha\beta}F^{\gamma\delta}F_{\gamma\delta}\right),
\end{equation}
where $\rho$, $\hat{p}$, $\xi$ and $u_{\alpha}$ are the energy
density, isotropic pressure, coefficient of bulk viscosity and
four-velocity associated with the fluid, respectively,
$F_{\alpha\beta}$ is the electromagnetic field tensor. We can write
the above equation with $p=\hat{p}-\xi\Theta$ as
\begin{equation}\label{3a}
T^-_{\alpha\beta}=(\rho+p)u_{\alpha}u_{\beta}+pg_{\alpha\beta}
-\eta{\sigma_{{\alpha}{\beta}}}+\frac{1}{4\pi}\left(F^\gamma_{\alpha}F_{\beta\gamma}
-\frac{1}{4}g_{\alpha\beta}F^{\gamma\delta}F_{\gamma\delta}\right),
\end{equation}
The four velocity in the comoving coordinates is
\begin{equation}\label{4}
u^{\alpha}=A^{-1}\delta^{\alpha}_{0},\quad u^{\alpha}u_{\alpha}=-1.
\end{equation}
Here $\eta>0$ is the coefficient of shearing viscosity, while the
shear tensor $\sigma_{\alpha\beta}$ is defined as
\begin{equation}\label{5}
\sigma_{\alpha\beta}=u_{(\alpha;b)}+a_{(\alpha} u_{\beta)}
-\frac{1}{3}\Theta(g_{\alpha\beta} + u_\alpha u_\beta),
\end{equation}
where $a_\alpha=u_{\alpha;\beta}u^\beta$ is the four acceleration
and $\Theta=u^{\alpha}_{;\alpha}$ is the expansion scalar. The
corresponding non-zero components turns out
\begin{equation}\label{6}
a_{1}=\frac{A'}{A},\quad a^\alpha a_{\alpha}=(\frac{A'}{AB})^2,\quad
\Theta=\frac{1}{A}\left(\frac{\dot{B}}{B}
+2\frac{\dot{R}}{R}\right),
\end{equation}
where dot and prime represent differentiation with respect to $t$
and $r$, respectively. The non-vanishing components for the shear
tensor are
\begin{eqnarray}\label{7}
\sigma_{11}={2B^2}\sigma,\quad {\sin^{2}\theta}\sigma_{22}
={\sigma_{33}}=-{R^2}\sigma,\quad
\sigma^{\alpha\beta}\sigma_{\alpha\beta}=2\sigma^2,
\end{eqnarray}
where
\begin{equation}\label{8}
\sigma=\frac{1}{3A}\left(\frac{\dot{R}}{R}-\frac{\dot{B}}{B}\right).
\end{equation}

\subsection{The Einstein-Maxwell Field Equations}

In four-vector formalism, the Maxwell equations are
\begin{eqnarray}\label{9}
F_{\alpha\beta}=\phi_{\beta,\alpha}-\phi_{\alpha,\beta}, \quad
F^{\alpha\beta}_{;\beta}=4{\pi}J^{\alpha},
\end{eqnarray}
where $\phi_\alpha$ and $J^{\alpha}$ are the four potential and
current density vector, respectively. We consider the charge to be
at rest resulting no magnetic field present in this local coordinate
system. Thus $\phi_\alpha$ and $J^{\alpha}$ can be written as
follows
\begin{eqnarray}\label{10}
\phi_{\alpha}=\Phi{\delta^0_\alpha},\quad
J^{\alpha}={\zeta}u^{\alpha},
\end{eqnarray}
where $\zeta(t,r)$ and $\Phi(t,r)$ describe the charge density and
the scalar potential, respectively. From Eqs.(\ref{9}) and
(\ref{10}), the only non-vanishing component of electromagnetic
field tensor is
\begin{eqnarray}\label{10a}
F_{10}=-F_{01}=\frac{\partial{\Phi}}{\partial{r}}.
\end{eqnarray}
The corresponding Maxwell equations are
\begin{eqnarray}\label{11}
\frac{\partial^2{\Phi}}{\partial{r^2}}-\left(\frac{A'}{A}+\frac{B'}{B}
-2\frac{R'}{R}\right)\frac{\partial{\Phi}}{\partial{r}}
&=&4\pi\zeta{AB^{2}},\\\label{12}
\frac{\partial^2{\Phi}}{\partial{t}\partial{r}}-\left(\frac{\dot{A}}{A}+\frac{\dot{B}}{B}
-2\frac{\dot{R}}{R}\right)\frac{\partial{\Phi}}{\partial{r}}&=&0.
\end{eqnarray}
Solving the above equations, it follows that
\begin{eqnarray}\label{13}
\frac{\partial{\Phi}}{\partial{r}}=\frac{qAB}{R^{2}},
\end{eqnarray}
where $q(r)$ is the total amount of charge from center to the
boundary surface of the star
\begin{equation}\label{14}
q(r)=4\pi\int^r_{0}{\zeta}B{R^2}dr.
\end{equation}
The electric field intensity is the charge per unit surface area of
the sphere
\begin{eqnarray}\label{15}
E(t,r)=\frac{q}{4{\pi}R^{2}}.
\end{eqnarray}

Using Eqs.(\ref{1}) and (\ref{3a}), we have non-vanishing components
of the Einstein-Maxwell field equations for (\ref{1}) yields
\begin{eqnarray}\nonumber
8{\pi}A^{2}({\rho}+2{\pi}E^2)&=&\left(\frac{2\dot{B}}{B}
+\frac{\dot{R}}{R}\right)\frac{\dot{R}}{R}-\left(\frac{A}{B}\right)^2
\\\label{16}&\times&\left[\frac{2R''}{R}
+\left(\frac{R'}{R}\right)^2-\frac{2B'R'}{BR}
-\left(\frac{B}{R}\right)^2\right],\\\label{17}
0&=&-2\left(\frac{\dot{R'}}{R}-\frac{\dot{R}A'}{RA}
-\frac{\dot{B}R'}{BR}\right), \\\nonumber
8{\pi}B^{2}(p+2\eta\sigma-2{\pi}E^2)&=&-\left(\frac{B}{A}\right)^2\left
[\frac{2\ddot{R}}{R}-\left(\frac{2\dot{A}}{A}
-\frac{\dot{R}}{R}\right) \frac{\dot{R}}{R}\right]\\\label{18}
&+&\left(\frac{2A'}{A}+\frac{R'}{R}\right)\frac{R'}{R}
-\left(\frac{B}{R}\right)^2,
\end{eqnarray}
\begin{eqnarray}\nonumber
8{\pi}R^{2}(p-\eta\sigma+2{\pi}E^2)&=&8{\pi}R^{2}(p-\eta\sigma
+2{\pi}E^2)\sin^{-2}\theta\\\nonumber
&=&-\left(\frac{R}{A}\right)^2\left[\frac{\ddot{B}}{B}
+\frac{\ddot{R}}{R}-\frac{\dot{A}}{A}
\left(\frac{\dot{B}}{B}+\frac{\dot{R}}{R}\right)
+\frac{\dot{B}\dot{R}}{BR}\right]\\\nonumber
&+&\left(\frac{R}{B}\right)^2\left[\frac{A''}{A}
+\frac{R''}{R}-\frac{A'}{A}\left(\frac{B'}{B}-\frac{R'}{R}\right)\right.
\left.-\frac{B'R'}{BR}\right].\\\label{19}
\end{eqnarray}
The Misner and Sharp \cite{28a} mass function $m(t,r)$ with charge
is given by
\begin{equation}\label{20}
m(t,r)=\frac{R}{2}(1-g^{\alpha\beta}R_{,\alpha}R_{,\beta})
=\frac{R}{2}\left(1+\frac{\dot{R}^2}{A^2}
-\frac{R'^2}{B^2}\right)+\frac{q^2}{2R}.
\end{equation}
The conservation equation, $(T^{-\alpha\beta})_{;\beta}=0$, yields
\begin{eqnarray}\label{21}
\dot{\rho}+(\rho+p)\left(\frac{\dot{B}}{B}+2\frac{\dot{R}}{R}\right)
+2{\eta\sigma}\left(\frac{\dot{B}}{B}-\frac{\dot{R}}{R}\right)=0,\\\label{22}
p'+2\eta{\sigma}'+(\rho+p)\frac{A'}{A}+2{\eta\sigma}
\left(\frac{A'}{A}+3\frac{R'}{R}\right)
-4{\pi}\frac{E}{R}(RE'+2R'E)=0.
\end{eqnarray}

\subsection{Junction Conditions}

We connect $M^{-}$ and $M^{+}$ metrics by considering the Darmois
junction conditions. For the smooth matching of these geometries, it
is required that the boundary is continuous and smooth. Thus the
continuity of first fundamental form of the metrics provides, i.e.,
\begin{equation}\label{23}
\left(ds^{2}_{-}\right)_{\Sigma}=\left(ds^{2}_{+}\right)_{\Sigma}
=\left(ds^{2}\right)_{\Sigma},
\end{equation}
and the continuity of second fundamental form of the extrinsic
curvature gives
\begin{equation}\label{24}
K^{+}_{ij}=K^{-}_{ij},\quad (i,j=0,2,3).
\end{equation}
Considering the interior and exterior spacetimes with Eq.(\ref{23}),
we have
\begin{eqnarray}\label{25}
\frac{dt}{d\tau}&=&A(t, r_{\Sigma})^{-1},\quad
R(t,r_{\Sigma})=r_{\Sigma}(\nu),\\\label{26}
\left(\frac{d\nu}{d\tau}\right)^{-2}&=&\left(1-\frac{2M}{r_{\Sigma}}
+\frac{Q^2}{r^2_{\Sigma}}+2\frac{dr_{\Sigma}}{d\nu}\right).
\end{eqnarray}
From Eq.(\ref{24}), the non-null components of the extrinsic
curvature turn out to be
\begin{eqnarray}\label{27}
&&K^{-}_{00}=-\left[\frac{A'}{AB}\right]_{\Sigma},\quad
K^{-}_{22}=\left[\frac{RR'}{B}\right]_{\Sigma},\quad
K^{-}_{33}=K^{-}_{22}\sin^{2}\theta, \\\label{28}
&&K^{+}_{00}=\left[\left(\frac{d^{2}\nu}{d\tau^{2}}\right)
\left(\frac{d\nu}{d\tau}\right)^{-1}
-\left(\frac{d\nu}{d\tau}\right)\left(\frac{M}{r^{2}}
-\frac{Q^{2}}{r^{3}}\right)\right]_{\Sigma}, \\\label{29}
&&K^{+}_{22}=\left[\left(\frac{d\nu}{d\tau}\right)
\left(1-\frac{2M}{r}-\frac{Q^2}{r^2}\right)r
+\left(\frac{dr}{d\tau}\right)r\right]_{\Sigma},\\\label{30}
&&K^{+}_{33}=K^{+}_{22}\sin^{2}\theta.
\end{eqnarray}
Using Eqs.(\ref{25})-(\ref{29}) and the field equations, we obtain
\begin{eqnarray}\label{31}
m\overset{\Sigma}=M,\quad p+2\eta{\sigma}\overset{\Sigma}=0.
\end{eqnarray}
where $q(r)=Q$ has been used. The above equation shows that across
the boundary, $\Sigma$, the masses of interior and exterior
spacetimes are matched as well as momentum flux is conserved.

\section{Perturbation Scheme and Collapse Equation}

Here we construct the collapse equation. For this purpose, we
perturb the field equations, dynamical equations and the mass
function upto first order in $\varepsilon$ by using the perturbation
scheme \cite{22,33}. We assume that initially all the physical
functions and the metric coefficients depend on $r$, i.e., the fluid
is unperturbed. Afterwards, all these quantities depend on time
coordinate, which are given by
\begin{eqnarray}\label{32}
A(t,r)&=&A_0(r)+\varepsilon T(t)a(r),\\\label{33}
B(t,r)&=&B_0(r)+\varepsilon T(t)b(r),\\\label{34}
R(t,r)&=&rB(t,r)[1+\varepsilon T(t)\bar{c}(r)],\\\label{35}
E(t,r)&=&E_0(r)+\varepsilon T(t)e(r),\\\label{36}
\rho(t,r)&=&\rho_0(r)+\varepsilon {\bar{\rho}}(t,r),
\end{eqnarray}
\begin{eqnarray}\label{37}
p(t,r)&=&p_{0}(r)+\varepsilon {\bar{p}}(t,r),\\\label{38}
\sigma(t,r)&=&\varepsilon {\bar{\sigma}}(t,r),\\\label{39}
m(t,r)&=&m_0(r)+\varepsilon{\bar{m}}(t,r),
\end{eqnarray}
where $0<\varepsilon\ll1$. For $\bar{c}=0$, we have shearfree
metric. Using this scheme, the static configuration of
Eqs.(\ref{16})-(\ref{19}) are written as
\begin{eqnarray}\label{40}
8{\pi}\left(\rho_{0}+2{\pi}E^{2}_{0}\right)
&=&-\frac{1}{B_0^2}\left[2\frac{B_0''}{B_0}-\left(\frac{B_0'}{B_0}\right)^2
+\frac{4}{r}\frac{B_0'}{B_0}\right], \\\label{41}
8{\pi}\left(p_{0}-2{\pi}E^{2}_{0}\right)
&=&\frac{1}{B_0^2}\left[\left(\frac{B_0'}{B_0}\right)^2
+\frac{2}{r}\left(\frac{A_0'}{A_0}+\frac{B_0'}{B_0}\right)
+2\frac{A_0'}{A_0}\frac{B_0'}{B_0}\right],\\\label{42}
8{\pi}\left(p_0+2{\pi}E^{2}_{0}\right)
&=&\frac{1}{B_0^2}\left[\frac{A_0''}{A_0}+\frac{B_0''}{B_0}
+\frac{1}{r}\left(\frac{A_0'}{A_0}+\frac{B_0'}{B_0}\right)
-\left(\frac{B_0'}{B_0}\right)^2\right].
\end{eqnarray}

The corresponding perturbed quantities upto first order in
$\varepsilon$ with Eqs.(\ref{32})-(\ref{38}) become
\begin{eqnarray}\nonumber
8{\pi}{\bar\rho}+32{\pi}^2{E_0}Te&=&-\frac{T}{B_0^3}
\left\{b\left(\frac{B_0'}{B_0}\right)^2-2b'
\left(\frac{B_0'}{B_0}-\frac{2}{r}\right)
+2b''+2B_0\right.\\\nonumber
&\times&\left.\left[\bar{c}''+\bar{c}'\left(2\frac{B_0'}{B_0}+\frac{3}{r}\right)
+\left(\frac{\bar{c}}{r^2}\right)
\right]\right\}\\\label{43}&-&24{\pi}\frac{Tb}{B_{0}}
\left({\rho_{0}}+2{\pi}E^{2}_{0}\right), \\\nonumber
0&=&2\left[\left(\frac{b}{A_0B_0}\right)'
+\left(\frac{\bar{c}}{A_0}\right)'+\left(\frac{1}{r}+\frac{B'_0}{B_0}\right)
\left(\frac{\bar{c}}{A_0}\right)\right]\dot{T},\\\label{44}\\\nonumber
8{\pi}({\bar{p}}+2\eta\bar{\sigma})-32{\pi}^2{E_0}Te&=&-\frac{2b}{A_0^2B_0}\ddot{T}+
\frac{2T}{B_0^2}\left[\left(\frac{1}{r}+\frac{B'_0}{B_0}\right)
\left(\frac{a}{B_0}\right)'\right.\\\nonumber
&+&\left.\left(\frac{b}{B_0}+\bar{c}\right)'\left(\frac{A_0'}{A_0}
+\frac{1}{r}+\frac{B'_0}{B_0}\right)
+\frac{\bar{c}}{r^2}\right.\\\label{45}&-&\left.
\bar{c}\left(\frac{B_0}{A_0}\right)^2\frac{\ddot{T}}{T}\right]
-16{\pi}\frac{Tb}{B_0}\left({p_{0}}-2{\pi}E^{2}_{0}\right),
\end{eqnarray}
\begin{eqnarray}\nonumber
8{\pi}({\bar{p}}-\eta\bar{\sigma})+32{\pi}^2{E_0}Te&=&-\frac{\ddot{T}}{A_0^2}
\left[2\frac{b}{B_0}+\bar{c}\right]+\frac{T}{B_0^2}
\left[\frac{a''}{A_0}+\frac{b''}{B_0}+\bar{c}'' \right.\\\nonumber
&+&\left.\frac{A''_0}{A_0}\left(\frac{a}{A_0}-\frac{b}{B_0}\right)+
\frac{b'}{B_0}\left(\frac{1}{r}-\frac{B_0'}{B_0}\right)
\right.\\\nonumber&+&\left.\frac{B'_0}{B_0}
\left(\bar{c}-\frac{b}{B_0}\right)'
+\frac{A_0'}{A_0}\left(\frac{b}{rB_0}+\bar{c}'\right)
\right.\\\label{46}&+&\left.
\frac{1}{r}\left(\frac{a}{A_0}+2\bar{c}\right)'\right]
-24{\pi}\frac{Tb}{B_0}\left({p_0}+2{\pi}E^{2}_{0}\right).
\end{eqnarray}

The static configuration for the dynamical equation (\ref{21}) is
trivially satisfied, while Eq.(\ref{22}) implies that
\begin{eqnarray}\label{47}
&&p'_0+(\rho_0+p_0)\frac{A_0'}{A_0}-4\pi{E_0}
\left[E'_{0}+2E_0\left(\frac{1}{r}
+\frac{B_0'}{B_0}\right)\right]=0,
\end{eqnarray}
which yields
\begin{eqnarray}\label{48}
\frac{A'_0}{A_0}&=&-\frac{1}{\rho_0+p_{0}}
\left\{p'_{0}-4\pi{E_0}\left[E'_{0}+2E_0\left(\frac{1}{r}
+\frac{B_0'}{B_0}\right)\right]\right\}.
\end{eqnarray}
The perturbed part of Eq.(\ref{21}) becomes
\begin{eqnarray}\label{49} &&\dot{\bar{\rho}}+(\rho_0+p_{r0})
\left(3\frac{b}{B_0}+2\bar{c}\right)\dot{T}=0,
\end{eqnarray}
which on integration leads to
\begin{eqnarray}\label{50}
\bar\rho=-(\rho_0+p_{0})\left(3\frac{b}{B_0}+2\bar{c}\right)T.
\end{eqnarray}
For the second dynamical equation, we have
\begin{eqnarray}\nonumber
&&\bar{p'}+2\eta{\bar{\sigma}}'+(\bar{\rho}+\bar{p})\frac{A'_0}{A_0}
+(\rho_0+p_{r0}){T}\left(\frac{a}{A_0}\right)'
\\\nonumber&+&2\eta{\sigma}\left(\frac{A_0'}{A_0}+\frac{3}{r}
+\frac{3B_0'}{B_0}\right)-4{\pi}T
\left[(eE_0)'+4e{E_0}\left(\frac{1}{r}
+\frac{B_0'}{B_0}\right)\right.\\\label{51}
&+&\left.2E^2_0\left(\frac{b}{B_0}\right)'+2E^2_0\bar{c}'\right]=0.
\end{eqnarray}

The unperturbed and perturbed configurations of electromagnetic
field are given by
\begin{eqnarray}\label{51a}
E_0(r)=\frac{q}{r^2B^2_0}, \quad
e(r)=-\left(\frac{2b}{B_0}+\bar{c}\right)E_0.
\end{eqnarray}
This shows that $e(r)$ depends upon the static configuration of
electromagnetic field. Similarly, we can write the static and
perturbed configuration for Eq.(\ref{20}) as
\begin{eqnarray}\label{52}
m_0&=&-\frac{rB_0}{2}\left(2r\frac{B'_0}{B_0}
+\left(r\frac{B'_0}{B_0}\right)^2\right)
+8\pi^2r^3{E^2_0}B^3_0,\\\nonumber
\bar{m}&=&-\left[r^2b'+r^3\frac{B'^2_0}{2B_0}
\left(2\frac{b'}{B'_0}-\frac{b}{B_0}\right)
\right]T-\bar{c}'r^3B_0\left(\frac{1}{r}
+\frac{B'_0}{B_0}\right)T\\\label{53}
&-&\bar{c}r^2B_0\left[\frac{3}{2}r
\left(\frac{B'_0}{B_0}\right)^2+3\frac{B'_0}{B_0}+\frac{1}{r}\right]T
+24\pi^2r^3B^2E^2_0\left(b+B_0\bar{c}\right)T.
\end{eqnarray}
From Eq.(\ref{52}), it follows that
\begin{equation}\label{54}
\frac{B_0'}{B_0}=-\frac{1}{r}+\frac{1}{r}
\sqrt{1-\frac{2m_0}{rB_0}+16{\pi}^2r^2E^2_0{B^2_0}}
\end{equation}
The perturbed configuration for the shear scalar is written as
\begin{equation}\label{55}
\bar{\sigma}=\frac{1}{3A_0}\bar{c}\dot{T}.
\end{equation}
The junction condition (\ref{31}) with Eqs.(\ref{37}) and (\ref{38})
leads to
\begin{equation}\label{56}
p_{0}\overset{\Sigma}=0,\quad
\bar{p}+2\eta{\bar{\sigma}}\overset{\Sigma}=0.
\end{equation}
Substituting the above relations in Eq.(\ref{45}), we get temporal
equation
\begin{equation}\label{57}
\ddot{T}(t)-\psi(r){T}(t)\overset{\Sigma}=0,
\end{equation}
where
\begin{eqnarray}\nonumber
\psi(r)&\overset{\Sigma}=&\left(\frac{A_{0}}{B_{0}}\right)^2
\left(\frac{b}{B_0}+\bar{c}\right)^{-1}
\left[\left(\frac{A_0'}{A_0}+\frac{1}{r} +\frac{B_0'}{B_0}\right)
\left(\frac{b}{B_0}+\bar{c}\right)'\right.\\\label{58} &+&\left.
\left(\frac{1}{r} +\frac{B_0'}{B_0}\right)
\left(\frac{a}{A_0}\right)'+\frac{\bar{c}}{r^2}
+16{\pi}^2E_0B_0(eB_0+bE_0)\right].
\end{eqnarray}
The general solution of Eq.(\ref{57}) yields
\begin{equation}\label{58a}
T(t)=c_1\exp(\sqrt{\psi_{\Sigma}}t)+c_2\exp(-\sqrt{\psi_{\Sigma}}t),
\end{equation}
where $c_1$ and $c_2$ are arbitrary constants. This provides two
independent solutions. Here we take $\psi_\Sigma$ to be positive for
physically meaningful result, i.e., when the system is in static
position, it starts collapsing at $t=-\infty$ when $T(-\infty)=0$
and goes on collapsing diminishing its areal radius with the
increase of $t$. The corresponding solution is found for $c_2=0$,
also, we choose $c_1=-1$, hence
\begin{equation}\label{59}
T(t)=-\exp(\sqrt{\psi_{\Sigma}}t).
\end{equation}

Next, we are interested to find instability range of the collapsing
fluid in terms of adiabatic index $\Gamma$. Chan et al. \cite{16}
found a relationship between $\bar{p}$ and $\bar{\rho}$ with an
equation of state of Harrison-Wheeler type \cite{34a} for the static
configuration as
\begin{eqnarray}\label{60}
\bar{p}=\Gamma\frac{p_0}{\rho_0+p_0}\bar{\rho},
\end{eqnarray}
where $\Gamma$ describes the change in pressure for a given change
in density (taken to be constant in the whole fluid distribution).
The above equation with Eq.(\ref{50}) leads to
\begin{eqnarray}\label{61}
\bar{p}=-{p_0}\Gamma\left(\frac{3b}{B_0}+2\bar{c}\right)T.
\end{eqnarray}
Using the above equation and (\ref{56}) in Eq.(\ref{45}), it follows
that
\begin{eqnarray}\nonumber
\left(\frac{a}{A_0}\right)'&=&B_0\left(\frac{1}{r}+\frac{B'_0}{B_0}\right)^{-1}
\left[-4\pi\left(\frac{3b}{B_0}+2\bar{c}\right)\Gamma{p_0}B_0
+8\pi{p_0}b-\frac{\bar{c}}{r^2B_0}\right.\\\nonumber
&-&\left.16\pi^2E_0(eB_0+bE_0)-\frac{1}{B_0}
\left(\frac{A_0'}{A_0}+\frac{1}{r}
+\frac{B_0'}{B_0}\right)\left(\frac{b}{B_0}+\bar{c}\right)'
\right.\\\label{62}&+&\left.\frac{B_0}{A^2_0}
\left(\frac{b}{B_0}+\bar{c}\right)\frac{\ddot{T}}{T}
+8\pi{\eta}B_0\frac{\bar{\sigma}}{T}\right].
\end{eqnarray}
We can write the collapse equation by substituting Eqs.(\ref{50}),
(\ref{55}), (\ref{61}) and (\ref{62}) in the perturbed configuration
of the second dynamical equation as follows
\begin{eqnarray}\nonumber
&&-\left[{p_0}\Gamma\left(\frac{3b}{B_0}+2\bar{c}\right)\right]'T
+\frac{2}{3A_0}\left(\eta{\bar{c}'}+\eta{\bar{c}}\frac{A'_0}{A_0}\right)
\dot{T}-\left(\frac{3b}{B_0}+2\bar{c}\right)T
\end{eqnarray}
\begin{eqnarray}\nonumber
&\times&\left(\rho_0+p_0(1+\Gamma)\right)\frac{A'_0}{A_0}
+(\rho_0+p_0)TB_0\left(\frac{1}{r}+\frac{B'_0}{B_0}\right)^{-1}
\left[\frac{B_0}{A^2_0}\left(\frac{b}{B_0}+\bar{c}\right)\frac{\ddot{T}}{T}
\right.\\\nonumber
&+&\left.8\pi{p_0}b+8\pi{\eta}B_0\frac{\bar{c}\dot{T}}{3A_0T}-
16\pi^2E_0(eB_0+bE_0)-4\pi\left(\frac{3b}{B_0}
+2\bar{c}\right)\Gamma{p_0}B_0 \right.\\\nonumber
&-&\left.\frac{1}{B_0} \left(\frac{A_0'}{A_0}+\frac{1}{r}
+\frac{B_0'}{B_0}\right)\left(\frac{b}{B_0}+\bar{c}\right)'
-\frac{\bar{c}}{r^2B_0}\right]+\frac{2}{3A_0}
\left(\frac{A_0'}{A_0}+\frac{1}{r}
+\frac{B_0'}{B_0}\right)\\\nonumber
&\times&\eta{\bar{c}}\dot{T}
-T\left[2E^2_0\left(\frac{b}{B_0}\right)'
+(eE_0)'+2E^2_0\bar{c}'+4\pi{E_0}\left(\frac{1}{r}
+\frac{B_0'}{B_0}\right)\right]=0.\\\label{63}
\end{eqnarray}

\section{Dynamical Instability of Charged Viscous Perturbation}

This section deals with the dynamical instability of charged viscous
fluid in the frame work of Newtonian and pN regimes. We see from
Eq.(\ref{34}) that the effects of shear explicitly appear only on
metric function $R(t,r)$. Using this fact, we can split
Eq.(\ref{44}) in two types of terms one with shearing viscosity and
other without it. Thus, the first term as well as the second and
third terms of Eq.(\ref{44}) are taken identically equal to zero
\begin{eqnarray}\label{64}
\left(\frac{b}{A_0B_0}\right)'=0,\quad
\left(\frac{\bar{c}}{A_0}\right)'
+\left(\frac{1}{r}+\frac{B'_0}{B_0}\right)
\left(\frac{\bar{c}}{A_0}\right)=0.
\end{eqnarray}
The first of the above equation provides
\begin{eqnarray}\label{65}
b=A_0B_0,
\end{eqnarray}
while the second leads to
\begin{eqnarray}\nonumber
\bar{c}'&=&-\bar{c}\left\{\left[\frac{p'_0}{\rho_0+p_0}
+\frac{4\pi{E_0}}{\rho_0+p_0}\left(E'_0+\frac{2E_0}{r}
\sqrt{1-\frac{2m_0}{rB_0}
+16{\pi}^2r^2E^2_0{B^2_0}}\right)\right]\right.\\\label{66}
&+&\left.\frac{1}{r}
\sqrt{1-\frac{2m_0}{rB_0}+16{\pi}^2r^2E^2_0{B^2_0}}\right\}.
\end{eqnarray}
Inserting Eqs.(\ref{48}), (\ref{54}), (\ref{59}), (\ref{65}) and
(\ref{66}) in the collapse equation, we have
\begin{eqnarray}\nonumber
&&-(3A_0+2\bar{c})\Gamma{p'_0}+\Gamma{p_0}\left\{A_0
\left[\frac{3p'_0}{\rho_0+p_0}
-\frac{12\pi{E_0}}{\rho_0+p_0}\left(E'_0+\frac{2E_0}{r}
\right.\right.\right.\\\nonumber &+&\left.\left.\left.
\sqrt{1-\frac{2m_0}{rB_0}+16{\pi}^2r^2E^2_0{B^2_0}}\right)\right]
+\frac{2p'_0\bar{c}}{\rho_0+p_0}
+\frac{2\bar{c}}{r}\sqrt{1-\frac{2m_0}{rB_0}+16{\pi}^2r^2E^2_0{B^2_0}}\right.\\\nonumber
&-&\left.\frac{4\pi{E_0}}{\rho_0+p_0}\left(E'_0+\frac{2E_0}{r}
\sqrt{1-\frac{2m_0}{rB_0}+16{\pi}^2r^2E^2_0{B^2_0}}\right)\right\}
-(\rho_0+p_0+\Gamma{p_0})\\\nonumber &\times&(3A_0+2\bar{c})
\left[\frac{p'_0}{\rho_0+p_0}-\frac{4\pi{E_0}}{\rho_0+p_0}\left(E'_0+\frac{2E_0}{r}
\sqrt{1-\frac{2m_0}{rB_0}+16{\pi}^2r^2E^2_0{B^2_0}}\right)\right]\\\nonumber
&+&(\rho_0+p_0)B_0\left(\frac{1}{r}
\sqrt{1-\frac{2m_0}{rB_0}+16{\pi}^2r^2E^2_0{B^2_0}}\right)^{-1}
\left\{\frac{8\pi}{3}\frac{B_0}{A_0}\eta{\bar{c}}\sqrt{\psi_{\Sigma}}\right.\\\nonumber
&-&\left.16{\pi}^2E_0(eB_0+bE_0)
+8\pi{p_0}A_0B_0-12\pi\Gamma{p_0}A_0B_0
+\left(\frac{B_0}{A_0}\right)\psi_{\Sigma}\right.\\\nonumber
&+&\left.\left(\frac{A_0}{B_0}\right)\left[\frac{1}{r}\sqrt{1-\frac{2m_0}{rB_0}
+16{\pi}^2r^2E^2_0{B^2_0}}
-\frac{p'_0}{\rho_0+p_0}+\frac{4\pi{E_0}}{\rho_0+p_0}
\left(E'_0+\frac{2E_0}{r}\right.\right.\right.\\\nonumber
&\times&\left.\left.\left.
\sqrt{1-\frac{2m_0}{rB_0}+16{\pi}^2r^2E^2_0{B^2_0}}\right)\right]
\left[\frac{p'_0}{\rho_0+p_0}-\frac{4\pi{E_0}}{\rho_0+p_0}\left(E'_0+\frac{2E_0}{r}
\right.\right.\right. \\\nonumber
&\times&\left.\left.\left.
\sqrt{1-\frac{2m_0}{rB_0}+16{\pi}^2r^2E^2_0{B^2_0}}\right)\right]
+\frac{\bar{c}}{B_0}\left[\frac{p'_0}{\rho_0+p_0}
-\frac{4\pi{E_0}}{\rho_0+p_0}\left(E'_0+\frac{2E_0}{r}
\right.\right.\right.\\\nonumber &\times&\left.\left.\left.
\sqrt{1-\frac{2m_0}{rB_0}+16{\pi}^2r^2E^2_0{B^2_0}}\right)\right]^2
-\frac{\bar{c}}{B_0}\left(\frac{2m_0}{r^3B_0}-16{\pi}^2E^2_0{B^2_0}
+8\pi{B^2_0}p_0\Gamma \right.\right.\\\nonumber &-&\left.\left.
\left(\frac{B_0}{A_0}\right)^2\psi_{\Sigma}\right)\right\}
+\frac{2}{3A_0r}\eta{\bar{c}}\sqrt{\psi_{\Sigma}}
\left[2\sqrt{1-\frac{2m_0}{rB_0}
+16{\pi}^2r^2E^2_0{B^2_0}}-\frac{p'_0}{\rho_0+p_0}\right.\\\nonumber
&+&\left.\frac{4\pi{E_0}}{\rho_0+p_0}\left(E'_0+\frac{2E_0}{r}
\sqrt{1-\frac{2m_0}{rB_0}+16{\pi}^2r^2E^2_0{B^2_0}}\right)\right]-
\left\{(eE_0)'+2(A_0+\bar{c})\right.\\\nonumber
&\times&\left.\left[-\frac{p'_0}{\rho_0+p_0}+\frac{4\pi{E_0}}{\rho_0+p_0}
\left(E'_0+\frac{2E_0}{r}\sqrt{1-\frac{2m_0}{rB_0}
+16{\pi}^2r^2E^2_0{B^2_0}}\right)\right]E^2_0\right.\\\label{67}
&+&\left.\frac{(4\pi{E_0}+2E^2_0\bar{c})}{r}\sqrt{1-\frac{2m_0}{rB_0}
+16{\pi}^2r^2E^2_0{B^2_0}}\right\}=0.
\end{eqnarray}

\subsection{Newtonian Limit}

Now we explore the instability range by applying the Newtonian
limit, i.e., $A_0=1,~B_0=1$ and ${\rho_0}\gg {p_0}$ and discarding
the terms of the order $\frac{m_0}{r}$ in Eq.(\ref{67}). It follows
that
\begin{eqnarray}\nonumber
&&-3p'_0\Gamma+4p'_0+\frac{(1+8\pi^2r^2E^2_0)}{r}\left(\frac{4}{3}\eta{\bar{c}}
\sqrt{\psi_{\Sigma}}-32\pi{E^2_0}-4\pi{E_0}\right)
\\\label{68}&-&16\pi{E_0E'_0}+(1-16\pi^2r^2E^2_0)r\rho_0{\psi_{\Sigma}}=0.
\end{eqnarray}
We see from Eq.(\ref{55}) that the perturbed configuration of the
shear scalar depends on the velocity gradients. Also, it is known
that the velocity of the particles for a self-gravitating star
increases towards the center. This shows that $\bar{\sigma}$ has
negative value inside the body and zero on the boundary surface. It
is worth noticing that for a collapsing body $\dot{T}<0$ as
$t{\rightarrow}-\infty$, thus Eq.(\ref{55}) implies that
$\bar{c}>0$. Considering the physical requirement, i.e., $p'_0<0$
and neglecting the terms like $\frac{\rho_0}{p_0}$ being
relativistic terms, we have instability condition (independent of
linear perturbation functions) for the charged viscous fluid as
\begin{eqnarray}\label{69}
\Gamma<\frac{4}{3}-\frac{4}{9}\eta{|{\bar{c}}|}
\frac{\sqrt{\psi_{\Sigma}}}{|{p'_0}|{r}}
+\frac{16\pi{E_0E'_0}}{3|{p'_0}|}+\frac{1+8\pi^2{r^2E^2_0}}{3|{p'_0}|{r}}
(32\pi{E^2_0}+4\pi{E_0}).
\end{eqnarray}
Here the critical value $\frac{4}{3}$ corresponds to the spherical
geometry and Newtonian gravity. In fact, $4$ in the numerator
corresponding to the weight of the envelop in Newtonian mechanics
varying as $r^{-2}$ which is distributed over the surface of sphere
yielding another $r^{-2}$. The denominator $3$ corresponds to the
volume of the sphere $r^3$. We note from the above equation that the
electromagnetic field diminishes the impact of shearing viscosity on
the dynamical instability and makes the system unstable at Newtonian
approximation. We retain the Newtonian classical result, i.e.,
$\Gamma<\frac{4}{3}$, for the case of shearfree or when fluid is not
charged viscous.

\subsection{Post-Newtonian Limit}

In view of pN limit, we consider
$A_0=1-\frac{m_0}{r},~B_0=1+\frac{m_0}{r}$ and the relativistic
corrections of the order $\frac{m_0}{r}$ in Eq.(\ref{67}) so that we
have
\begin{eqnarray}\nonumber
&-&(3+2\bar{c})p'_0\Gamma+(4+2\bar{c})p'_0+\frac{(1+8\pi^2r^2E^2_0)}{r}
\left(\frac{4}{3}\eta{\bar{c}}\sqrt{\psi_{\Sigma}}+4\pi{E_0}
\right.\\\nonumber
&+&\left.2E^2_0\bar{c}-8\pi{E^2_0}(2+2\bar{c})\right)
+(1-16\pi^2r^2E^2_0)r\rho_0[8\pi{p_0}+\psi_{\Sigma}
\end{eqnarray}
\begin{eqnarray}\nonumber
&-&16\pi^2{E_0}(e+E_0)+\frac{8\pi}{3}\eta{\bar{c}}\sqrt{\psi_{\Sigma}}]
-(4+2\bar{c})4\pi{E_0E'_0}-(eE_0)'\\\label{70}&+&16\pi^2r\bar{c}\rho_0E^2_0
+\rho_0r\bar{c}\psi_{\Sigma}(1-8\pi^2r^2E^2_0)=0.
\end{eqnarray}
Here we apply the same procedure as in the Newtonian limit and
ignoring terms with higher order $\frac{m_0}{r}$, we get the
instability range at pN limit as follows
\begin{eqnarray}\nonumber
&&\Gamma<\frac{4}{3}-\frac{1}{3|{p'_0}|}
\left[\left(\frac{4}{3r}\eta{|{\bar{c}}|}\sqrt{\psi_{\Sigma}}+4\pi{E_0}
+2E^2_0|{\bar{c}}|\right)(1+8\pi^2r^2E^2_0)\right.\\\nonumber
&+&\left.16\pi^2E^2_0r\rho_0(|{\bar{c}}|+16\pi^2{E_0}r^2(e+E_0))
+2|\bar{c}|+\rho_0{\psi_{\Sigma}}r(1+|{\bar{c}}|) \right.\\\nonumber
&+&\left.\frac{8\pi}{3}
\eta{\rho_0}r{|{\bar{c}}|}\sqrt{\psi_{\Sigma}}\right]
+\frac{16\pi^2rE_0\rho_0}{3|{p'_0}|}
\left[\frac{8\pi}{3}\eta{|{\bar{c}}|}r^2E_0\sqrt{\psi_{\Sigma}}+e+E_0
\right.\\\nonumber
&+&\left.r^2E_0\psi_{\Sigma}(1+\frac{|{\bar{c}}|}{2})\right]
+\frac{1}{3|{p'_0}|}\left[8\pi{\rho_0}p_0r+4\pi{E_0E'_0}(4+2|{\bar{c}}|)+(eE_0)'
\right.\\\label{71} &+&\left.16\pi{E^2_0}(1+|{\bar{c}}|)
\frac{(1+8\pi^2r^2E^2_0)}{r}\right].
\end{eqnarray}
This equation shows that the positive terms diminish the
relativistic effects of negative terms occurring from the shearing
viscosity and electromagnetic field at pN approximation, making the
fluid more unstable. It is mentioned here that instability condition
depends upon the static configuration of the system, as perturbed
variable $e(r)$ and $b(r)$ depend on static configuration given in
Eqs.(\ref{51a}) and (\ref{65}) respectively.

\section{Conclusion}

We have investigated the role of electromagnetic field on the
instability conditions (\ref{69}) and (\ref{71}) at Newtonian and pN
approximations. Pinheiro and Chan \cite{35} found that the
collapsing stars with huge amount of charge
($Q\sim5.408\times10^{20}$ Coulomb) leads to the
Reissner-Nordstr$\ddot{o}$m black hole. They concluded that the
models with charge to mass ratio $\frac{Q}{m_0}\leq0.631$ would form
a black hole. Ernesto and Simeone \cite{36} examined the stability
of charged thin shell and found that Reissner-Nordstr$\ddot{o}$m
geometry for different values of charge either have an inner and
outer event horizon or a naked singularity. This shows the relevance
of charge on the evolution and structure formation of astrophysical
objects.

In general, the behavior of electromagnetic field is always positive
being the Coulomb's repulsion force. Sharif and Abbas \cite{1}
explored that due to the weak nature of the electromagnetic field,
the end state of collapsing cylinder results a charged black string.
We see from Eq.(\ref{69}) that shearing viscosity boost the
stability of the fluid which is the consequence of the fact that the
collapse with shear proceeds faster than without shear. Whereas the
electromagnetic field being a positive quantity diminishes the
effects of shearing viscosity and makes the fluid unstable at
Newtonian approximation. This corresponds to the fact that charge
delays the event horizon formation or even halts the complete
contraction of the star \cite{10}. Also, the electromagnetic field
has the same impact for Eq.(\ref{71}) at the pN approximation. It is
worth mentioning here that our results for electromagnetic field are
consistent with the results obtained in \cite{33}.

Finally, we would like to mention that we have made all the
discussion on the hypersurface, where the areal radius is constant.
The solution of the temporal equation (60) includes oscillating and
non-oscillating functions correspond to stable and unstable systems.
For the sake of instability conditions, we have confined our
interest in the non-oscillating ones. We have investigated the role
of physical quantities in the onset of dynamical instability of
fluid during the collapse, hence the instability conditions contain
those terms which have radial dependence.

\vspace{0.5cm}

{\bf Acknowledgments}

\vspace{0.5cm}

We would like to thank the Higher Education Commission, Islamabad,
Pakistan, for its financial support through the {\it Indigenous
Ph.D. 5000 Fellowship Program Batch-VII}. One of us (MA) would like
to thank University of Education, Lahore for the study leave.

\end{document}